\documentclass{sig-alternate}

\usepackage[latin1]{inputenc}
\usepackage{epsfig, epic, eepic, amscd}

\begin{document}

\conferenceinfo{ISSAC'07,} {July 29--August 1, 2007, Waterloo, Ontario, Canada.}

\CopyrightYear{2007}

\crdata{978-1-59593-743-8/07/0007}


\newcommand{\calA}{{\cal A}}
\newcommand{\calB}{{\cal B}}
\newcommand{\calF}{{\cal F}}
\newcommand{\calG}{{\cal G}}
\newcommand{\calR}{{\cal R}}
\newcommand{\calZ}{{\cal Z}}
\def\D{\mathcal{D}}
\def\L{\mathcal{L}}
\def\S{\mathcal{S}}
\def\I{\mathcal{I}}
\def\V{\mathcal{V}}
\def\E{\mathcal{E}}

\newcommand{\N}{{\mathbb N}}
\newcommand{\Z}{{\mathbb Z}}
\newcommand{\Q}{{\mathbb Q}}
\newcommand{\R}{{\mathbb R}}
\newcommand{\C}{{\mathbb C}}
\newcommand{\K}{{\mathbb K}}
\newcommand{\kk}{{\mathrm k}}

\def\J{\mathrm{J}}
\def\x{\mathrm{x}}
\def\a{\mathrm{a}}
\def\d{\mathrm{d}}

\def\GL{\mathrm{GL}}
\def\det{\mathrm{det}}
\def\SL{\mathrm{SL}}
\def\PSL{\mathrm{PSL}}
\def\PGL{\mathrm{PGL}}
\def\O{\mathrm{O}}

\def\gl{\mathfrak{gl}}
\def\g{\mathfrak{g}}
\def\h{\mathfrak{h}}

\newcommand{\Frac}[2]{\displaystyle \frac{#1}{#2}}
\newcommand{\Sum}[2]{\displaystyle{\sum_{#1}^{#2}}}
\newcommand{\Prod}[2]{\displaystyle{\prod_{#1}^{#2}}}
\newcommand{\Int}[2]{\displaystyle{\int_{#1}^{#2}}}
\newcommand{\Lim}[1]{\displaystyle{\lim_{#1}\ }}

\newenvironment{menumerate}{%
    \renewcommand{\theenumi}{\roman{enumi}}%
    \renewcommand{\labelenumi}{\rm(\theenumi)}%
    \begin{enumerate}} {\end{enumerate}}
     
\newenvironment{system}[1][]%
	{\begin{eqnarray} #1 \left\{ \begin{array}{lll}}%
	{\end{array} \right. \end{eqnarray}}

\newenvironment{meqnarray}%
	{\begin{eqnarray}  \begin{array}{rcl}}%
	{\end{array}  \end{eqnarray}}

\newenvironment{marray}%
	{\\ \begin{tabular}{ll}}
	{\end{tabular}\\}

\newenvironment{Pmatrix}
        {$ \left( \!\! \begin{array}{rr} } 
        {\end{array} \!\! \right) $}

\newtheorem{example}{Example}
\newtheorem{theorem}{Theorem}
\newtheorem{lemma}{Lemma}
\newtheorem{corollary}{Corollary}
\newtheorem{definition}{Definition}
\newtheorem{proposition}{Proposition}
\newtheorem{remark}{Remark}

\newcommand{\fleche}[3]{#1 \stackrel{#2}\longrightarrow #3}
\def\ssi{si et seulement si\ }
\newcommand{\tab}{\hspace*{\fill}}
\newcommand{\bs}{{\backslash}}
\newcommand{\eps}{{\varepsilon}}
\newcommand{\into}{{\;\rightarrow\;}}
\newcommand{\PD}[2]{\frac{\partial #1}{\partial #2}}
\def\Hat{\widehat}
\def\Bar{\overline}
\def\vect{\vec}
\def\fbar{{\bar f}}
\def\xbar{{\bar \x}}
\newcommand{\afaire}{$$\vdots$$ \begin{center} {\sc a faire ...} \end{center} $$\vdots$$ }
\newcommand{\pref}[1]{(\ref{#1})}

\def\Maple{{\sc Maple}}
\def\RG{{\sc Rosenfeld-Gr\"obner}}



\newcommand{\algf}{\sffamily}
\newcommand{\BEGIN}{{\algf begin}}
\newcommand{\END}{{\algf end}}
\newcommand{\IF}{{\algf if}}
\newcommand{\THEN}{{\algf then}}
\newcommand{\ELSE}{{\algf else}}
\newcommand{\ELIF}{{\algf elif}}
\newcommand{\FI}{{\algf fi}}
\newcommand{\WHILE}{{\algf while}}
\newcommand{\FOR}{{\algf for}}
\newcommand{\DO}{{\algf do}}
\newcommand{\OD}{{\algf od}}
\newcommand{\RETURN}{{\algf return}}
\newcommand{\PROCEDURE}{{\algf procedure}}
\newcommand{\FUNCTION}{{\algf function}}
\newcommand{\INDENTER}{{\algf si} \=\+\kill}

\newcommand{\target}{\mathop{\mathrm{t}}}
\newcommand{\source}{\mathop{\mathrm{s}}}
\newcommand{\trdeg}{\mathop{\mathrm{tr~deg}}}
\newcommand{\jet}[2]{\jmath_{#1}^{#2}}
\newcommand{\rank}{\operatorname{rank}}
\newcommand{\sign}{\operatorname{sign}}
\newcommand{\ord}{\operatorname{ord}}
\newcommand{\aut}{\operatorname{aut}}
\newcommand{\Hom}{\operatorname{Hom}}
\newcommand{\codim}{\operatorname{codim}}
\newcommand{\coker}{\operatorname{coker}}
\newcommand{\rp}{\operatorname{rp}}
\newcommand{\leader}{\operatorname{ld}}
\newcommand{\card}{\operatorname{card}}
\newcommand{\Fr}{\operatorname{Frac}}
\newcommand{\NF}{\operatorname{\mathsf{normal\_form}}}
\newcommand{\rang}{\operatorname{rang}}

\def \Id{\mathrm{Id}}

\def \diff{\mathrm{Diff}^{\mathrm{loc}} }
\def \diffg{\mathrm{Diff} }
\def \Esc{\mathrm{Esc}}

\newcommand{\initial}{\mathop{\mathsf{init}}}
\newcommand{\separant}{\mathop{\mathsf{sep}}}
\newcommand{\rem}{\mathop{\mathsf{rem}}}
\newcommand{\quo}{\mathop{\mathsf{quo}}}
\newcommand{\pquo}{\mathop{\mathsf{pquo}}}
\newcommand{\lcoeff}{\mathop{\mathsf{lcoeff}}}
\newcommand{\mvar}{\mathop{\mathsf{mvar}}}

\newcommand{\prem}{\mathop{\mathsf{prem}}}
\newcommand{\remp}{\mathrel{\mathsf{partial\_rem}}}
\newcommand{\remf}{\mathrel{\mathsf{full\_rem}}}
\renewcommand{\gcd}{\mathop{\mathrm{gcd}}}
\renewcommand{\deg}{\mathop{\mathrm{deg}}}
\newcommand{\pairs}{\mathop{\mathrm{pairs}}}
\newcommand{\dd}{\mathrm{d}}
\newcommand{\ideal}[1]{(#1)}
\newcommand{\cont}{\mathop{\mathrm{cont}}}
\newcommand{\pp}{\mathop{\mathrm{pp}}}
\newcommand{\pgcd}{\mathop{\mathrm{pgcd}}}
\newcommand{\ppmc}{\mathop{\mathrm{ppcm}}}
\newcommand{\init}{\mathop{\mathrm{initial}}}

\title{Towards a New ODE Solver Based on Cartan's Equivalence Method}
%
%
%
%
%

\numberofauthors{2} 
%
\author{
%
%
\alignauthor
Raouf Dridi\\
       \affaddr{Laboratoire d'Informatique Fondamentale de Lille}\\
       \affaddr{Bureau 334 B\^atiment M3}\\
       \affaddr{ 59655 Villeneuve d'Ascq CEDEX - FRANCE }\\
       \email{dridi@lifl.fr} 
\alignauthor
Michel Petitot\\
       \affaddr{Laboratoire d'Informatique Fondamentale de Lille}\\
       \affaddr{Bureau 334 B\^atiment M3}\\
       \affaddr{ 59655 Villeneuve d'Ascq CEDEX - FRANCE }\\
       \email{petitot@lifl.fr}
}

\date{\today}

\maketitle
\begin{abstract}
	The aim of the present paper is to propose an algorithm for a new 
	ODE--solver which should improve the abilities of current  solvers to handle second order
	differential equations. The paper provides also a theoretical result revealing the relationship
	between the change of coordinates, that maps the generic equation 
	to a given target equation,  and  the symmetry $\D$-groupoid of this target.
\end{abstract} 

\category{I.1.2H.4}{Symbolic and Algebraic  Manipulation}{Computing Methodologies}[Algorithms]

\terms{Algorithms}


\section{Introduction}
\begin{figure}[!hbt]\label{introfig} 
\setlength{\unitlength}{0.00083333in}
\begingroup\makeatletter\ifx\SetFigFont\undefined%
\gdef\SetFigFont#1#2#3#4#5{%
  \reset@font\fontsize{#1}{#2pt}%
  \fontfamily{#3}\fontseries{#4}\fontshape{#5}%
  \selectfont}%
\fi\endgroup%
{\renewcommand{\dashlinestretch}{30}
\begin{picture}(3851,3799)(0,-10)
\put(1875,885){\ellipse{2162}{656}}
\put(1787,2649){\ellipse{1324}{408}}
\path(1780,1599)(1780,1230)
\path(1765.275,1288.900)(1780.000,1230.000)(1794.725,1288.900)
\path(1817,539)(1817,256)(1144,256)
\path(1200.540,270.135)(1144.000,256.000)(1200.540,241.865)
\path(1817,256)(2558,256)
\path(2501.460,241.865)(2558.000,256.000)(2501.460,270.135)
\thicklines
\put(2603,161){\arc{90}{1.5708}{3.1416}}
\put(2603,352){\arc{90}{3.1416}{4.7124}}
\put(3220,352){\arc{90}{4.7124}{6.2832}}
\put(3220,161){\arc{90}{0}{1.5708}}
\path(2558,161)(2558,352)
\path(2603,397)(3220,397)
\path(3265,352)(3265,161)
\path(3220,116)(2603,116)
\put(796,3297){\arc{90}{1.5708}{3.1416}}
\put(796,3695){\arc{90}{3.1416}{4.7124}}
\put(2773,3695){\arc{90}{4.7124}{6.2832}}
\put(2773,3297){\arc{90}{0}{1.5708}}
\path(751,3297)(751,3695)
\path(796,3740)(2773,3740)
\path(2818,3695)(2818,3297)
\path(2773,3252)(796,3252)
\thinlines
\path(1780,3259)(1780,2843)
\path(1765.000,2901.500)(1780.000,2843.000)(1795.000,2901.500)
\thicklines
\put(104,104){\arc{120}{1.5708}{3.1416}}
\put(104,359){\arc{120}{3.1416}{4.7124}}
\put(1066,359){\arc{120}{4.7124}{6.2832}}
\put(1066,104){\arc{120}{0}{1.5708}}
\path(44,104)(44,359)
\path(104,419)(1066,419)
\path(1126,359)(1126,104)
\path(1066,44)(104,44)
\put(2896,2564){\arc{90}{1.5708}{3.1416}}
\put(2896,2820){\arc{90}{3.1416}{4.7124}}
\put(3762,2820){\arc{90}{4.7124}{6.2832}}
\put(3762,2564){\arc{90}{0}{1.5708}}
\path(2851,2564)(2851,2820)
\path(2896,2865)(3762,2865)
\path(3807,2820)(3807,2564)
\path(3762,2519)(2896,2519)
\thinlines
\path(2476,2669)(2815,2669)
\path(2768.500,2657.000)(2815.000,2669.000)(2768.500,2681.000)
\path(1801,2444)(1801,2039)
\path(1786.000,2084.000)(1801.000,2039.000)(1816.000,2084.000)
\thicklines
\put(1246,1664){\arc{90}{1.5708}{3.1416}}
\put(1246,1968){\arc{90}{3.1416}{4.7124}}
\put(2421,1968){\arc{90}{4.7124}{6.2832}}
\put(2421,1664){\arc{90}{0}{1.5708}}
\path(1201,1664)(1201,1968)
\path(1246,2013)(2421,2013)
\path(2466,1968)(2466,1664)
\path(2421,1619)(1246,1619)
\put(1604,3543){\makebox(0,0)[lb]{{\SetFigFont{8}{9.6}{\rmdefault}{\bfdefault}{\updefault}Input :}}}
\put(1958,2200){\makebox(0,0)[lb]{{\SetFigFont{7}{8.4}{\rmdefault}{\mddefault}{\itdefault}No}}}
\put(2806,220){\makebox(0,0)[lb]{{\SetFigFont{8}{9.6}{\rmdefault}{\bfdefault}{\updefault}Fail}}}
\put(2064,362){\makebox(0,0)[lb]{{\SetFigFont{7}{8.4}{\rmdefault}{\mddefault}{\itdefault}No}}}
\put(1497,362){\makebox(0,0)[lb]{{\SetFigFont{7}{8.4}{\rmdefault}{\mddefault}{\itdefault}Yes}}}
\put(2551,2744){\makebox(0,0)[lb]{{\SetFigFont{7}{8.4}{\rmdefault}{\mddefault}{\itdefault}Yes}}}
\put(1126,944){\makebox(0,0)[lb]{{\SetFigFont{8}{9.6}{\rmdefault}{\mddefault}{\itdefault} Does the equation admit}}}
\put(826,3344){\makebox(0,0)[lb]{{\SetFigFont{8}{9.6}{\rmdefault}{\mddefault}{\updefault} Ordinary differential equation  }}}
\put(226,119){\makebox(0,0)[lb]{{\SetFigFont{8}{9.6}{\rmdefault}{\bfdefault}{\updefault}of the order}}}
\put(226,269){\makebox(0,0)[lb]{{\SetFigFont{8}{9.6}{\rmdefault}{\bfdefault}{\updefault}Reduction }}}
\put(2926,2594){\makebox(0,0)[lb]{{\SetFigFont{8}{9.6}{\rmdefault}{\bfdefault}{\updefault}Integration  }}}
\put(1351,1694){\makebox(0,0)[lb]{{\SetFigFont{8}{9.6}{\rmdefault}{\bfdefault}{\updefault}Lie equations }}}
\put(1351,1844){\makebox(0,0)[lb]{{\SetFigFont{8}{9.6}{\rmdefault}{\bfdefault}{\updefault}Resolution of }}}
\put(1276,2594){\makebox(0,0)[lb]{{\SetFigFont{8}{9.6}{\rmdefault}{\mddefault}{\itdefault}Known equation?     }}}
\put(826,794){\makebox(0,0)[lb]{{\SetFigFont{8}{9.6}{\rmdefault}{\mddefault}{\itdefault} 1-parameter symmetry group?     }}}
\end{picture}
}
    \caption{{\it  \small {General flowchart of typical ODE-solver.}}}
        \label{introfig0}
\end{figure}
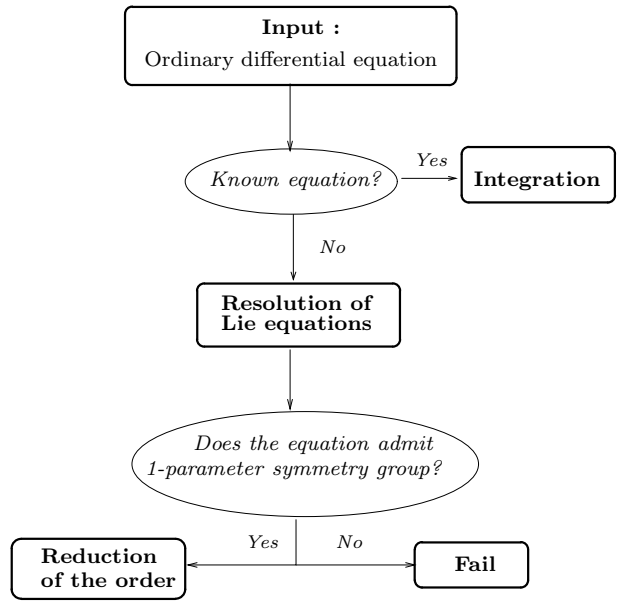

Current ODE-solvers make use of a 
combination of symmetry methods and classification methods. 
Classification methods are used when the ODE matches
a recognizable pattern (that is, for which a solving method is
already implemented), and symmetry methods are reserved for
the non-classifiable cases -- Fig. \ref{introfig0}.
Using symmetry methods, the solvers first look for the generators
of 1-parameter symmetry  groups of the given ODE, and then use  this
information to integrate it, or at least reduce its 
order \cite{Terrab97, Terrab98}.

In practice,  present solvers are  often unable to return closed form solution.
Consider for instance  the  following  equation
\begin{equation}\label{ODE1}
 y'' =-{y}^{3}{y'}^{4}  -{\frac {{y'}^{2}}{y}}-\frac{1}{2}y,
\end{equation}
which   admits  only one  1-parameter symmetry  group. 
Using this information, actual solvers return a complicated first order ODE
and a quadrature. Clearly, such output is quite useless for practical applications. 
More dramatically, consider the following  equation 
\begin{equation}\label{ODE2}
y'' = {\frac {2\,{x}^{4}y'-6\,{y}^{2}x-1}{{x}^{5}}}.
\end{equation}
When applied actual  solvers, output no result. This failure is due to the fact that the
above equation does not match any recognizable pattern and has zerodimensional point symmetry
group(oid). Thus neither  symmetry  method nor classification method works.

Our solver (the implementation  is in progress) is designed to handle
 such equations.  It returns an equation from  
Kamke's book \cite{Kamke}, equivalent to the equation to be solved, and the equivalence 
transformation~$\varphi$. Thus, for the equation  \pref{ODE1}
 we obtain the Rayleigh equation $y'' + {y'}^4 + y=0$ (number 72 in \cite{Kamke}) and the change  of coordinates 
$
	\varphi: (x,y) \into (x, \,y^2/2).
$ 
For the equation \pref{ODE2}, we obtain the  first Painlevé equation  
$y'' =6y^2+x$  (number 3 in \cite{Kamke}) and the change of coordinates 
$
	\varphi: (x,y) \into (1/{x},\,  y).
$
It is  worth noticing that this transformation  can be
 composed with the point symmetries of  the Painlevé  equation given by
$
     (x,y) \into (\lambda^2 x,\, \lambda y)
$ 
 with $\lambda^5=1$.

To summarize the theoretical result of the paper, let $\E_f$ denote the generic  ODE $y''=f(x,y,y')$ and
$\Phi$ an allowed  $\D$-groupoid acting on the variables $x$ and $y$. Suppose
that $\Phi$ is given by quasi-linear Lie defining equations.
Define  $\S_f := \aut(\E_f)\cap\Phi$ where $\aut(\E_f)$ denotes
the (local) contact symmetry  $\D$-groupoid of $\E_f$. 
Given   $\fbar\in\Q\left(x,y,y' \right)$ and  assume that the symmetry
$\D$-groupoid $\S_\fbar$ is zerodimensional. The transformation $\varphi\in\Phi$ mapping the
generic equation  $\E_f$ to the target equation $\E_\fbar$ is called the 
\emph{necessary form of the change of coordinates}. We shall see that this transformation
 exists and belongs to the differential field
$\overline{\Q\langle f\rangle}$, endowed with the partial derivations $\left(\partial_x,\, \partial_y,\partial_{y'}\right)$,
for almost any analytic function $f$ satisfying
$\varphi_*(\E_f)=\E_\fbar$. In other words, $\varphi$ is an algebraic
function in $f$ and its partial derivatives and can be obtained without solving differential
equations. Moreover, the degree of $\varphi$ is equal to  $\card(\S_\fbar)$ which is finite.
Note that, the use of the $\D$-groupoids formalism is
dictated by the non global  invertibility of the transformation $\varphi$. 

As we shall see,  $\varphi$ can be
obtained using differential elimination. Unfortunately, such approach is rarely effective due to expressions
swell. For this reason, we propose in section \ref{cartan} a new method
to precompute the transformation $\varphi$ in terms of differential
invariants, for each target equation $\E_\fbar$ in Kamke's list. These
invariants are provided by Cartan's  method. In the last
section, we present our solver. This solver uses 7 possible types of
transformations
$\Phi_1,\cdots, \Phi_7$. Using Lie infinitesimal method we
precalculate to each target equation a signature. That is, the dimensions
of the 7 symmetry groupoids associated to the 7 groupoids $\Phi_1,\cdots,
\Phi_7$. If two differential equations are equivalent then their
signatures \emph{match}. This fact significantly restricts the space of research in 
kamke's list at the run-time (when the input equation $\E_f$ is known).

\section{Equivalence problems}

\subsection{Groupoids}
\begin{definition}[Groupoid]
	A \emph{groupoid} is a category in which every arrow is invertible.
\end{definition}
Let $(\calG,\circ,\source,\target)$ be a category.
Each arrow $\varphi \in \calG$ admits a source $\source(\varphi)$ and a
target $\target(\varphi)$  which are \emph{objects} of this category. For each arrows 
$\alpha,\ \beta \in \calG$ such that  $\source(\beta) = \target(\alpha)$, 
there exists a unique arrow $\beta \circ \alpha \in \calG$ with the  source  $\source(\alpha)$ 
and the target  $\target(\beta)$. 
If  $\calG$  is a groupoid, for each arrow $\varphi \in \calG$,
there exists a unique inverse arrow $\varphi^{-1}$ such that 
$\varphi^{-1}\circ\varphi=\Id_{\source(\varphi)}$ and 
$\varphi \circ \varphi^{-1}=\Id_{\target(\varphi)}$.

Let $X$ and $U$ be two manifolds and $\x \in X$.
The Taylor series up to order $q$ (i.e. the jet of order $q$) of a
function  $f:X \to U$, of 
 class $C^q$,  is denoted $\jet \x q f$. 
The Taylor series  of~$f$ about  $\x$ is denoted $\jet \x {} f$ or $\jet \x \infty f$.
We shall say that  $\x \in X$ is the source and  $f(\x)\in U$ 
is the target of the $q$-jet $\jet x q f$. 

\begin{example} 
	For instance, when  $X=U=\C$, we have
	$$
	\jet x q f := \left( x, f(x), f'(x), \ldots, f^{(q)}(x) \right) \in \C^{q+2}.
	$$
	This jet is said to be \emph{invertible}
	if $f'(x) \neq 0$. 
	The jet of the function $\Id$ about the  point $ x$ is $(x,x,1,0,\ldots, 0)$.
\end{example} 

For each  $q \in \N$ and each $\x \in X$, we set 
$\J_\x^q (X,U) := \bigcup_f \jet \x q f$ and $\J^q (X,U) := \bigsqcup_{\x \in X}\J_\x^q (X,U)$.
We denote by $\J_*^q(X,X)$ the submanifold of $\J^q(X,X)$ formed by the  invertible jets.
 $\J_*^q(X,X)$ is a groupoid \cite{OlverGroupoid} for the composition of Taylor series up to order
$q$ according to 
\begin{equation} \label{def:comp}
	\jet{\x}{q} (g \circ f) = \left( \jet{f(\x)}{q} g \right) \circ 
			\left( \jet{\x}{q} f \right).
\end{equation}

By definition, a  $\D$-groupoid \cite{Malgrange} $\calG \subset \J_*^\infty(X,X)$
is a sub--groupoid of $\J_*^\infty(X,X)$ formed by the Taylor series 
solutions (see def. \ref{TS:solns}) of an algebraic PDE system called 
the \emph{Lie defining equations}. This  system 
contains an inequation which  expresses   the  invertibility of  the jets.  
The set of smooth functions $\varphi:X\to X$ that are local solutions of the Lie
defining equations of $\calG$ is a \emph{pseudo-group} denoted by $\Gamma \calG$. We define
$\dim \calG := \dim C$ and, if $\dim C = 0$, $\deg \calG := \deg C$  where~$C$ is a characteristic
set (see sect. \ref{diffalg}) of the Lie defining equations. 
We have $\deg \calG = \card(\Gamma \calG)$.

\begin{example} [$\Phi_3$]
   Let $\Phi_3$ (see  table 1) be the $\D$-groupoid of infinite jets of transformations
   $(\bar x, \bar y) = \varphi(x,y)$ where
   \begin{equation} \label{def:Phi3}
       \bar x = x + C \mbox{ and } \bar y = \eta(x,y).
   \end{equation}
   The constant $C \in \C$  and the function $\eta:\C^2\to\C$ are arbitrary.
   $\Phi_3\subset \J_*^\infty(\C^2,\C^2)$ is an infinite dimensional $\D$-groupoid where the corresponding Lie defining equations are  
   \begin{equation} \label{Phi3:eqns}
       {\bar x}_x = 1,\  {\bar x}_y = 0,\ {\bar y}_y \neq 0.
   \end{equation}
\end{example}

\begin{definition}[$\calG$-Invariant] \label{def:invariant}
   An \emph{invariant} of the  $\D$-groupoid $\calG \subset \J_*^\infty(X,X)$ is a function 
   $I:X \to \C$ which is constant on the orbits of~$\calG$.
\end{definition}
Clearly, the sum, the product and the ratio of two invariant functions is still 
an invariant function. Consequently, invariant functions of  $\calG$ define a field.

\subsection{Differential equations and diffieties}

Let $\E_f$ denotes the generic ODE
\begin{equation} \label{def:yn}
	y^{(n+1)}=f(x,\,y,\,y',\ldots,\,y^{(n)}).
\end{equation}
Let $M :=\J^n(\C,\C)$ be the $n$--th order jets space \cite{Olver1} of functions
from $\C$ to $\C$. Let 
$
	\x := (x,\,y,\,y_1,\,\ldots,\,y_n)\in \C^{m}
$
be a local coordinates system over $M$ where $m:=n+2=\dim M$.

Every  differential equation $\E_f$ defines  a \emph{diffiety}
\cite{Vinogradov}. This diffiety is given by  the 
manifold $M$  and a set of 1-forms, called  \emph{contact forms}, satisfying the
Frobenius condition of complete integrability. Contact forms are linear combinations of the basic 
contact 1-forms $\d y-y_1 \d x,\, \d y_1-y_2 \d x, \cdots,$\, $\d y_n-f(\x) \d x$. 
Vector fields which are orthogonal to the contact forms are colinear to the \emph{Cartan field}
\begin{equation*}
	D_x := \PD{}{x} + y_1 \PD{}{y} + y_2 \PD{}{y_1} + \cdots + f(\x) \PD{}{y_n}.
\end{equation*}
They generate a  distribution denoted by $\Delta_f$. 
A \emph{local isomorphism} $\varphi$ between two diffieties $\E_f$ and $\E_\fbar$ is, 
by definition, a local  diffeomorphism $\varphi:M \to M$ such that
\begin{equation*}
	\Delta_\fbar = \varphi_*(\Delta_f).
\end{equation*}

\subsection{Equivalence problem and symmetries}

\begin{definition}[Equivalence problem]
	An \emph{equivalence problem (EPB)} is  an ordered pair $(M, \Phi)$ where $M = \J^n(\C,\C)$ and
	$\Phi \subset \J_*^\infty(\C^2,\,\C^2)$ is a $\D$-groupoid of
	point transformations from $\C^2$ to  $\C^2$.
\end{definition}
There exists a unique prolongation of $\Phi$, denoted  $\Phi^{(n)}$, that acts on $M$ (see section \ref{prolong}).
Two differential equations $\E_f$ and $\E_\fbar$ are said to be
equivalent if there exists a local transformation $\varphi:M \to M$ satisfying 
the differential system
\begin{equation}\label{PDEsystem}
         \Delta_\fbar = \varphi_*(\Delta_f) \mbox{ and } \varphi\in \Gamma\Phi^{(n)}.
\end{equation}
The second condition means that $\varphi$ fulfills the Lie defining
equations of the $\D$-groupoid $\Phi^{(n)}$.

The system \pref{PDEsystem} is fundamental and we shall see that it 
can be treated by two different approaches : brute-force method based  on 
differential algebra (section \ref{diffalg}) and geometric approach relying  
on Cartan's theory of exterior  differential systems (section \ref{cartan}). 
It is classically known  that the existence of at least one transformation $\varphi$ 
can be checked by  computing  the \emph{integrability conditions} of the system
\pref{PDEsystem}, which is completely algorithmic whenever  $f$ and $\fbar$ are 
explicitly known \cite{cartan:pb, Olver1, Boulier}. 
However, there is no general algorithm for computing closed form of $\varphi$.
In the sequel, we shall show that if the  function  $\fbar$ is  fixed 
such that a certain   
$\D$-groupoid $\S_\fbar$ is zerodimensional, then $\varphi$ is obtained \emph{without} 
integrating any differential equation.

\begin{definition}[$\S_\fbar$]
	To any   EPB,  with fixed target equation $\E_\fbar$, we associate 
	the $\D$-groupoid $\S_\fbar \subset \J_*^\infty(M,M)$ formed
	by  the Taylor series 
	solutions  of the \emph{self--equivalence} problem
	\begin{equation}\label{S:eqns}
         	\Delta_\fbar = \sigma_*(\Delta_\fbar) \mbox{ and } \sigma \in \Gamma \Phi^{(n)}.
	\end{equation}
\end{definition}

\begin{example} 
	Consider the EPB $(\J^1(\C,\C), \Phi_3)$ and
	the Emden-Fowler equation $\E_\fbar$ (number 11 in \cite{Kamke}) 
	\begin{equation}\label{emf:eq}
		y''= \Frac{1}{x y^2}.
	\end{equation}

	The Lie  defining equations of the $\D$-groupoid $\S_\fbar$ are given by 
	the characteristic set
	\begin{equation}\label{symemf}
		\left\{ \bar p = \Frac{\bar yp}{y},\ {\bar y^3}=y^3,\ \bar x=x \right\}.
	\end{equation}
	This PDE system is particular. Indeed, it contains only non differential equations. 
	We have $\dim \S_\fbar = 0$ and $\deg \S_\fbar = 3$.
	We deduce that its associated pseudo-group 
	$$
		\Gamma \S_\fbar = \left\{ (x, y, p) \to (x, \lambda y, \lambda p)
			\mid \lambda^3 = 1 \right\}
	$$
	is, actually, a group with 3 elements.
\end{example}

\subsubsection*{Equivalence problem and associated $\D$-groupoid}

Let  $X := \J^\infty(M, \C)$. 
Any EPB $(M, \Phi)$ defines a $\D$-groupoid $\calG \subset \J^\infty_*(X,X)$ 
formed by the set of triplets
$$(\jet\x{}f,\, \jet\x{}\varphi,\, \jet{\varphi(\x)}{}\fbar)$$ 
where $\x \in M$ and the  functions $(f,\, \varphi,\, \fbar)$ are local 
solutions of the differential system \pref{PDEsystem}.
The source of a triplet is the infinite jet $\jet{\x}{}f \in X$ and 
the target is the infinite jet ${\jet{\varphi(\x)}{}\fbar \in X}$. 
The composition of two triplets $(\jet\x{}f,\, \jet\x{}\varphi_1,\, \jet{\x_1}{}f_1)$ 
and $(\jet{\x_1}{}f_1,\, \jet{\x_1}{}\varphi_2,\, \jet{\x_2}{}f_2)$
is the triplet $(\jet{\x}{}f,\, \jet{\x}{}\varphi,\, \jet{\x_2}{}f_2)$
 where we have ${\varphi := \varphi_2 \circ \varphi_1}$.


\begin{definition} [Specialized   invariant]\label{def:invspecial}
	For each $\calG$-invariant $I$ and each   function $f: M \to \C$, we define 
	the \emph{specialized   invariant} {${I[f]: M \to \C}$ by}
	\begin{equation} 
		I[f](\x) := I(\jet{\x}{} f),\quad \x \in M.
	\end{equation}
\end{definition}

\section{Using differential algebra} \label{diffalg}

The aim of this section is  to use differential elimination  to solve the EPB when the target function $\fbar$ is 
a  $\Q$-rational function, explicitly known  and  the $\D$-groupoid of
symmetries $\S_\fbar$ is zerodimensional.

\subsection{The vocabulary}
The reader is assumed to be familiar with the basic notions  and 
notations of  differential algebra. Reference books are
\cite{Ritt} and \cite{kolchin:livre}. We also refer to \cite{BLOP, hubert00, Boulier}.
Let $U=\{u_1,\cdots,u_n\}$ be a set of differential 
indeterminates. $\kk$ is a differential field of 
characteristic zero endowed with the set of derivations 
$\Delta=\left\{ \partial_1,\,\cdots,\partial_p \right\}$. The monoid of derivations
\begin{equation}
	\Theta := \left\{ \partial_1^{\alpha_1} \partial_2^{\alpha_2} \cdots
		\partial_p^{\alpha_p} \mid \alpha_1, \ldots, \alpha_p \in \N \right\}
\end{equation}
acts freely on the alphabet $U$ and defines a new (infinite) alphabet $\Theta U$.
The differential ring of the polynomials built over $\Theta U$ with 
coefficients in $k$ is denoted $R=\kk\{U\}$.
Fix an admissible ranking over $\Theta U$. For $f \in R$, 
$\leader(f) \in \Theta U $ denotes the \emph{leader} (main variable), 
$I_f \in R$ denotes the \emph{initial} of $f$ and $S_f\in R$ denotes
the separant of $f$. Recall that $S_f=\frac{\partial f}{\partial v}$ where $ v = \leader(f)$. 
Let $C\subset R$ be a finite set of differential polynomials. 
Denote by $[C]$ the differential ideal generated by $C$ and by $\sqrt{[C]}$ 
the radical of $[C]$. Let $H_C := \{ I_f \mid f \in C\} \cup \{ S_f \mid f \in C \}$.
As usual, $\remf$ is the Ritt full reduction algorithm \cite{kolchin:livre}. 
If $r = \remf(f,C)$ then $\exists h \in H^\infty_C,\, hf=r \mod [C]$. 
Then the \emph{normal form} is defined by $\NF(f) := r/h$.
\begin{definition}[Characteristic set]
	The set $C\subset R$ is said to be a \emph{characteristic set} of the differential ideal
	${\mathfrak{c} := \sqrt{[C]}:H_C^\infty}$ if  
	\begin{marray}
		(1)  & $C$ is autoreduced, \\
		(2)  & $f\in \mathfrak{c}$ if and only if $\remf(f,\, C) = 0$.
	\end{marray}
\end{definition}

\begin{definition}[Quasi--linear characteristic set]
	The  characteristic set $C\subset R$ is said to be
	\emph{quasi--linear} if for each $f\in C$
	we have $\deg(f,\,v) = 1$ where~$v$ is the leader of~$f$.
\end{definition}

\begin{proposition} \label{C:C0}
	When the characteristic set $C$ is quasi--linear, the
	differential ideal $\mathfrak{c} := \sqrt{[C]}:H_C^\infty \subset R$ is prime.	
\end{proposition}
	
\subsection{Taylor series solutions space}

Let $\kk :=\Q(x_1,\cdots,x_p)$ be the differential field of coefficients endowed with the set
of derivations 
$\left\{ \frac{\partial}{\partial x_1}, \cdots, \frac{\partial}{\partial x_p} \right\}$. 
Let $C$ be a characteristic
set of a prime differential ideal $\mathfrak{c} \subset R$.
We associate to $C$ the  system
\begin{equation} \label{C:PDEsyst}
	(C=0, H_C\neq 0)
\end{equation} 
of equations $f=0, \, f \in C$ and inequations $h \neq 0,\, h \in H_C$.
\begin{definition} [Taylor series solution]\label{TS:solns}
	A \emph{Taylor series solution} of the PDE system \pref{C:PDEsyst}
	above is a morphism $\mu: R\into \C$
	of (non differential) $\Q$-algebras  such that
	$$[C] \subset \ker \mu \mbox{ and } H_C \cap \ker \mu =\emptyset.$$
\end{definition}
The morphism $\mu$ defines an infinite jet where the  source is
$\source(\mu) := \left( \mu(x_1),\ldots, \mu(x_p) \right) \in \C^p$ and the target is
$\target(\mu) := \left( \mu(u_1),\ldots, \mu(u_n) \right) \in \C^n$.
Thus, a Taylor series is a $\C$-point
of an algebraic quasi--affine variety. Its Zarisky cloture is an affine variety 
defined by the ideal $\mathfrak{c}$. 
The dimension of the solutions space of \pref{C:PDEsyst} is
the number of arbitrary constants appearing in the Taylor series solutions $\mu$ when the
source point $\x := \source(\mu) \in \C^p$ is determined.
Let $K$ be the fractions field $\Fr (R/\mathfrak{c})$.
Recall that the \emph{transcendence degree} of a field extension $K/\kk$ is the  greatest number
of elements in $K$ which are $\kk$-algebraically independent. 
The degree $[K:\kk]$ is the dimension of $K$ as a
$\kk$-vector space. When $\trdeg (K/\kk)=0$, the field $K$ is algebraic over $\kk$ and ${[K:\kk] < \infty}$. 
If $f \in C$, we denote $\rank(f) := (v,d)$ where $v := \leader f$ and $d := \deg(f, v)$.
Let
\begin{eqnarray*}
	\rank C   &:=& \left\{ \rank(f)\mid f \in C \right\} \\
	\leader C &:=& \left\{ \leader(f)\mid f \in C \right\} \\
	\dim C &:=& \card \left( \Theta U \setminus \Theta(\leader C) \right) \\
	\deg C &:=& \Prod{f\in C}{} \deg(f, \leader f).
\end{eqnarray*}
\begin{proposition} \label{C:solns}
	$\dim C = \trdeg(K/\kk)$ is the dimension of the solutions space of \pref{C:PDEsyst}.
	If $\dim C = 0$ then the cardinal of the solutions space is finite
	and equal to $\deg C = [K:\kk]$.
\end{proposition}

\subsection{Differential elimination} \label{sect:elim}

Let $U=U_1 \sqcup U_2$ be a partition of the alphabet $U$.
A ranking which eliminates the indeterminates of $U_2$ is such that
\begin{equation}
	\forall v_1 \in \Theta U_1,\, \forall v_2 \in \Theta U_2, \quad  v_2 \succ v_1.
\end{equation}

Assume that $C$ is a characteristic set of 
the prime differential ideal $\mathfrak{c} = \sqrt{[C]}:H_C^\infty$ w.r.t.
the elimination ranking $\Theta U_2 \succ \Theta U_1$.
Let $R_1 := k\{ U_1 \}$ be the differential polynomials $k$-algebra generated by the set
$U_1$. Consider the set $C_1 := C \cap R_1$ and
the differential ideal $\mathfrak {c}_1 := \mathfrak{c} \cap R_1$.

\begin{proposition}\label{elim:prop}	
 	$C_1$ is a characteristic set of $\mathfrak{c}_1$.
\end{proposition}

Consider the differential field of fractions  $K := \Fr (R/\mathfrak{c})$ and
denote by $\alpha: R \to K$ the canonical $k$-algebra morphism. 
Let $K_1$ be the differential subfield of $K$ generated by the set $\alpha(R_1)$. 
Then $K_1$ is the fraction field associated to the prime differential ideal 
$\mathfrak{c}_1 := \mathfrak{c} \cap R_1$.
The partition of the  characteristic set
\begin{equation} \label{def:partition}
	C = C_1 \sqcup C_2 \quad (\mbox{i.e. } C_2 := C \setminus C_1).
\end{equation}
enables us to study the field extension $K/K_1$. 
\begin{proposition}
   	$\trdeg (K/K_1) = \dim C_2$. 
	If $\dim C_2 =0$ then $[K:K_1] = \deg C_2$.
\end{proposition}

\subsection{The system \pref{PDEsystem} revisited}
\subsubsection{Prolongation algorithm}\label{prolong}
 Our aim, here, is to prolong the  action of $\Phi\subset \J_*^\infty(\C^2, \C^2)$
 on the manifold  $M := \J^n(\C, \C)$. For each integer $q\geq 0$, define 
\begin{eqnarray*}
	\kk^{(q)} &:=& \Q(x,\, y,\, y_1,\ldots,\, y_q) \\
	R^{(q)}   &:=& \kk^{(q)}\{ \bar x,\, \bar y,\, \bar y_1, \ldots,\, \bar y_q \}
\end{eqnarray*}
The differential field $\kk^{(q)}$ is the coefficients field
of the ring of differential polynomials $R^{(q)}$
endowed with the set of derivations $\left\{ \frac{\partial}{\partial x},\, 
\frac{\partial}{\partial y},\,\ldots,\, \frac{\partial}{\partial y_q} \right\}$.
Let us assume that the  Lie defining equations of $\Phi$ are given by 
a characteristic set $C^{(0)} \subset R^{(0)}$. The  
$\D$-groupoid  $\Phi^{(q)}$  acting on  $\J^q$ and  prolonging the action of $\Phi$ 
 is characterized by a characteristic set $C^{(q)} \subset R^{(q)}$.
The prolongation formulae \cite{Olver} of the  point transformation 
$(x,y) \to \left( \xi(x,y),\, \eta(x,y) \right)$ are of the form
$$
	\bar y_q = \eta_q(x,\,y,\ldots,\, y_q), 
$$
where $\bar y = \eta(x,\, y)$ if $q=0$.
The computation of the characteristic set  $C^{(q)}$ is done incrementally using 
the infinite Cartan field $D_x := \PD{}{x} + y_1 \PD{}{y} + y_2 \PD{}{y_1} + \cdots$
\begin{eqnarray*}
        \eta_{q} &:=& D_x \eta_{q-1} \cdot {(D_x \xi)}^{-1} \\
        C^{(q)}  &:=& C^{(q-1)} \cup \left\{ \bar y_{q} - 
        	\NF\left( \eta_{q},\, C^{(q-1)}\right) \right\}
\end{eqnarray*}

\begin{proposition} \label{prop:prolongation}
	If $C^{(0)}$ is a \emph{quasi-linear} characteristic set of $\Phi$
	then  $C^{(q)}$ is a \emph{quasi-linear} characteristic set of $\Phi^{(q)}$
	w.r.t. the elimination ranking 
	$\Theta \bar y_q \succ \Theta \bar y_{q-1} \succ \cdots \succ \Theta\{\bar y, \bar x\}$.
\end{proposition}
The previous proposition gives an efficient method to prolong a 
$\D$-groupoid $\Phi$ without the explicit knowledge of transformations.

\subsubsection{EPB with fixed target}
Let us compute a characteristic set $C[\fbar] \subset R^{(n)}\{ f \}$
for the PDE system \pref{PDEsystem} where  $\fbar$ is fixed
$\Q$-rational function. First, prolong  $C^{(0)}$ up to the order
$n+1$ as above. Then
$C[\fbar]$ is obtained by substituting in $C^{(n+1)}$ the indeterminate 
$y_{n+1}$ by the  symbol $f$ and the indeterminate  $\bar y_{n+1}$
by the rational function  $\bar f(\bar x,\, \cdots,\, \bar y_n)$.

\begin{example}
   For the EPB $(\J^1(\C, \C), \Phi_3)$, we have
   \begin{equation} \label{Sigmaexple}
	\begin{gathered}
	\bar p - \bar y_x - p\bar y_y=0,\\
	\bar y_{xx} + 2p\bar y_{xy} + p^2\bar y_{yy} + f\ \bar y_y 
	-\fbar(\bar x,\,\bar y,\,\bar p) =0,  \\
	\bar x_x-1=0,\, \bar x_y=0,\, \bar x_p=0,\, \bar y_p=0,\, \bar y_y\neq 0.
	\end{gathered}
   \end{equation} 
   These equations constitute a quasi-linear characteristic set w.r.t. the elimination
   ranking $\Theta f \succ \Theta\bar p\succ \Theta\bar y\succ \Theta\bar x$. 
   Hence, the associated differential ideal is prime.
\end{example}

\begin{corollary} \label{cor:cible}	
	The PDE system \pref{PDEsystem} (where  $\fbar$ is
	a fixed $\Q$-rational function)
	is a quasi--linear characteristic set $C[\fbar] \subset R^{(n)}\{ f \}$
	w.r.t the elimination ranking 
	$\Theta f \succ \Theta \bar y_n \succ \cdots \succ \Theta\{\bar y, \bar x\}$.
\end{corollary}

\subsection{Brute-force method} \label{brute:force}
Using  \RG\ we compute  
a new characteristic set $C[\fbar]$ of the PDE system \pref{PDEsystem}
w.r.t. the new  ranking 
$\Theta\{ \bar y_n,\,\cdots,\,\bar y_1, \bar x \} \succ \Theta\{ f \}$.
We make the partition of $C := C[\fbar]$ as in  \pref{def:partition}
\begin{equation} \label{bf:partition}
	C = C_f \sqcup C_\varphi
\end{equation}
where $C_f := C \cap \kk^{(n)}\{f\}$ and 
$C_\varphi := C \setminus C_f$.

\begin{proposition}
	The transformation $\varphi$ does exist for \emph{almost any} function $f$ 
	satisfying the PDE system associated to the characteristic set $C_f[\fbar]$. 
	The function $\bar \x=\varphi(x)$ is solution of the
	PDE system associated to  $C_\varphi[\fbar]$. 
\end{proposition}
If $\dim C_\varphi[\fbar] = 0$, one can calculate $\varphi$ by an algebraic process without 
integrating differential equations.

\begin{definition}
	When $\dim C_\varphi[\fbar] = 0$, the  algebraic system associated to 
	$C_\varphi[\fbar]$ is called 
	the \emph{necessary form of the change of coordinates} $\xbar = \varphi(\x)$.
\end{definition}

\begin{example}
Consider the EPB $(\J^1(\C, \C), \Phi_3)$. Suppose that the target $\E_\fbar$ is the Airy 
equation 
$$
	\bar y''= \bar x \bar y.
$$
In this case, \RG\ returns $C_\varphi[\fbar]$ and $C_f[\fbar]$ resp. given by \pref{thairyetac} and  \pref{thairyf}
\begin{meqnarray}\label{thairyetac}
	\bar y_{{xx}} &=& -f \bar y_{{y}}+pf_{{p}}\bar y_{{y}}-1/2\,{p}^{2}f_{pp}
		\bar y_{{y}}+\bar y f_{{y}}-1/2\,\bar y f_{{xp}} \\
		&&-1/2\,\bar y f_{{pp}}f +1/4\,\bar y {f_{{p}}}^{2}-1/2\,\bar y pf_{{yp}}\\
	\bar y_{{xy}} &=& -1/2\,f_{{p}}\bar y_{{y}}+1/2\,pf_{{p,p}}\bar y_{{y}}\\
	\bar y_{{yy}} &=& -1/2\,f_{{pp}}\bar y_{{y}},\quad	\bar y_p = 0,\\
	\bar x &=& f_{{y}}-1/2\,f_{{xp}}-1/2\,f_{{pp}}f +1/4\,{f_{{p}}}^{2}-1/2\,pf_{{yp}}
\end{meqnarray}%
\begin{meqnarray} \label{thairyf}
	f_{{xxp}} &=& 2\,f_{{xy}}+f_{{p}}f_{{xp}}
		-2+{p}^{2}f_{{yyp}}-f_{{pp}}f_{{x}} +\cdots\\ 
	f_{{xyp}} &=& 2\,f_{{yy}}-pf_{{yyp}}-f_{{ypp}}f -f_{{pp}}f_{{y}}+f_{{p}}f_{{yp}}\\
	f_{{xpp}} &=& f_{{yp}}-pf_{{ypp}}\\
	f_{ppp}   &=& 0.
\end{meqnarray}%
We have $\dim C_\varphi[\fbar] = 3$ which means that the transformation 
$\xbar = \varphi(\x)$, when $f$ satisfies $C_f[\fbar]$, depends on 3 arbitrary constants.
\end{example}

\begin{example}
Consider the EPB $(\J^1(\C,\C), \Phi_1)$ 
where  $\Phi_1$ is defined in table 1. Assume that  
 the target equation  $\E_\fbar$ is
$$
	 \bar y''= \bar y^3.
$$
Here, \RG\ returns  $C_\varphi[\fbar]$  and $C_f[\fbar]$  resp.  given by   \pref{eq:ex3} and \pref{f:ex3}
\begin{meqnarray} \label{eq:ex3}
	\bar y^{2} &=& 1/12\,({4\,f_{{y}}-2\,f_{{xp}}-2\,f_{{pp}}f_{{}}-2\,pf_{{yp}}
		+{f_{{p}}}^{2}}),\\
	\bar x &=& x,
\end{meqnarray}%
\begin{meqnarray}\label{f:ex3}
	f_{{xxxp}} &=&({4\,f_{{y}}-2\,f_{{xp}}-2\,f_{{pp}}f_{{}}-2\,pf_{{yp}}
		+{f_{{p}}}^{2}})^{-1}\times\\
		&& \vdots\\
	f_{{xxyp}} &=& ({4\,f_{{y}}-2\,f_{{xp}}-2\,f_{{pp}}f_{{}}-2\,pf_{{yp}}
		+{f_{{p}}}^{2}})^{-1} \times\\
		&& \vdots\\
	f_{{xyyp}} &=& \,\,({4\,f_{{y}}-2\,f_{{xp}}-2\,f_{{pp}}f_{{}}-2\,pf_{{yp}}
		+{f_{{p}}}^{2}})^{-1} \times\\
		&& \vdots\\
	f_{{xpp}} &=& f_{{yp}}-pf_{{ypp}}\\
	f_{{ppp}} &=& 0.
\end{meqnarray}%
Consequently $\dim C_\varphi[\fbar] = 0$ and $\deg C_\varphi[\fbar] = 2$.
Thus, $\varphi$ is the algebraic transformation of degree 2, given by equations~\pref{eq:ex3}.
\end{example}

\subsection{Discrete symmetries $\D$-groupoids} \label{self:EPB}

The self--equivalence problem,  is in fact, the EPB when the  PDE system \pref{PDEsystem}
is \emph{specialized} by substituting the symbol $f$ by the value 
$\fbar(\x)$, that is 
\begin{equation} \label{S:spec}
	f := \fbar(x,\, y,\, \ldots,\, y_n).
\end{equation}
After specialization, the differential system $C_f[\fbar]$ constraining the function $f$
is automatically satisfied (since there exists at least one solution
$\xbar = \sigma(\x)$ of the problem, namely  $\sigma=\Id$). The symmetries $\sigma$ are 
solutions of a characteristic set $C_\sigma[\fbar]$ obtained form $C_\varphi[\fbar]$ by the 
specialization \pref{S:spec}.

By definition, the degree of an algebraic transformation $\bar \x=\varphi(x)$ is
the generic number of points $\xbar$ when $\x$ is determined.

\begin{theorem}\label{thm2}
	The following conditions are equivalent 
	\begin{marray}
		(1) & $\dim(C_\varphi[\fbar])=0$, \\
		(2) & $\dim(\S_\fbar)=0$, \\
		(3) & $\deg(\S_\fbar) < \infty$.
	\end{marray}%
	In this case, $\deg \S_\fbar = \deg(C_\varphi[\fbar]) = \deg \varphi$.
\end{theorem}

\begin{proof}
        Define 
        $$
        	\calG_\fbar := 
         	\left\{ (\jet\x{}f,\, \jet\x{}\varphi,\, \jet{\xbar}{}\fbar) \in \calG \mid 
         		\fbar \mbox{ determined}
         	\right\}.
        $$
	$\calG_{\fbar}$ is an algebraic covering of $M$ defined by the characteristic set
	$C_\varphi[\fbar]$.     
        The $\D$-groupoid $\S_\fbar \subset \calG_{\fbar}$ is defined by differential 
        system \pref{S:eqns} i.e. the characteristic set
        $C_\sigma[\fbar]$. Figure  \ref{groupoidefig} shows that
        $\S_\fbar$ acts simply transitively on $\calG_\fbar$.
\begin{figure}[h]
        \begin{center}
\setlength{\unitlength}{0.00027333in}%
\begingroup\makeatletter\ifx\SetFigFont\undefined%
\gdef\SetFigFont#1#2#3#4#5{%
  \reset@font\fontsize{#1}{#2pt}%
  \fontfamily{#3}\fontseries{#4}\fontshape{#5}%
  \selectfont}%
\fi\endgroup%
{\renewcommand{\dashlinestretch}{30}
\begin{picture}(3311,2233)(0,-10)
\drawline(421,1672)(2349,270)
\drawline(2281.998,297.054)(2349.000,270.000)(2302.612,325.401)
\drawline(2775,1774)(2775,289)
\drawline(2755.845,365.610)(2775.000,289.000)(2794.155,365.610)
\drawline(582,1924)(2491,1924)
\drawline(2419.960,1906.240)(2491.000,1924.000)(2419.960,1941.760)
\put(2700,1849){\makebox(0,0)[lb]{{\SetFigFont{12}{14.4}{\rmdefault}{\mddefault}{\updefault}
\small{$\jet{\xbar_0}{} \fbar$}
}}}
\put(-950,1849){\makebox(0,0)[lb]{{\SetFigFont{12}{14.4}{\rmdefault}{\mddefault}{\updefault}
\small{$\jet{\x}{} f$}
}}}
\put(1425,2074){\makebox(0,0)[lb]{{\SetFigFont{12}{14.4}{\rmdefault}{\mddefault}{\updefault}
\small{$\varphi_0$}
}}}
\put(2850,874){\makebox(0,0)[lb]{{\SetFigFont{12}{
14.4}{\rmdefault}{\mddefault}{\updefault}
\small{$\sigma$}
}}}
\put(595,874){\makebox(0,0)[lb]{{\SetFigFont{12}{14.4}{\rmdefault}{\mddefault}{\updefault}
\small{$\varphi$}
}}}
\put(2475,-240){\makebox(0,0)[lb]{{\SetFigFont{12}{14.4}{\rmdefault}{\mddefault}{\updefault}
\small{$\jet{\xbar}{} \fbar$}
}}}
\end{picture}
}
        	\caption {\it {Simply transitive action of $\S_\fbar$ on $\calG_\fbar$
        		where $\xbar_0 = \varphi_0(\x)$ and $\xbar=\varphi(\x)$}}
        	\label{groupoidefig}
        \end{center}
\end{figure}

   	Choose a  point $\xbar_0$ in $M$. For every $\varphi_0\in \Gamma \Phi$,
   	define the rational transformation $\S_\fbar \to \calG_\fbar$
   	$$
   		\jet{\xbar_0}{}\sigma \to 
   			(\jet{\xbar_0}{}\sigma)\circ (\jet{\x}{}\varphi_0),
   			\quad (\sigma \in \Gamma \S_\fbar).
   	$$   	
        In fact, according to the Taylor series composition formulae, this transformation is birational.
	Thus, the  one-to-one correspondence between the two algebraic varieties
	$\calG_{\fbar}$ and~$\S_\fbar$ is birational. Consequently,
	 the two characteristic sets $C_\varphi[\fbar]$ and
         $C_\sigma[\fbar]$ have  the  same
	dimension and the same degree.
\end{proof}

\begin{lemma}
	The rank of the characteristic set $C_{\varphi}[\fbar]$ is stable under
	the specialization \pref{S:spec} i.e. $\rank C_\varphi[\fbar] = \rank C_\sigma[\fbar]$.
\end{lemma}
\begin{proof}
	The specialization \pref{S:spec} transforms the
	characteristic set $C_\varphi[\fbar]$ to $C_\sigma[\fbar]$. 
	A fall of the rank of $C_\varphi[\fbar]$
	during the specialization contradicts the existence of
	birational correspondence between $\calG_{\fbar}$ and  $\S_\fbar$.
\end{proof}	

\begin{remark}
When the transformation $\xbar := \varphi(\x)$ is locally bijective but not globally,  
$\S_f$ and $\S_\fbar$ need not to have the same degree. Indeed, consider again the
groupoid $\Phi_3$ and the equations 
$$
	y''= \Frac{6y^4+x-2{y'}^2}{2y}  \mbox{ and } \bar y'' = 6\bar y^2 + \bar x
$$
which are equivalent under  $(\bar x=x, \  \bar y=y^2)$. The corresponding 
symmetry group are respectively given by 
$$
	\Gamma \S_f= \{ (x,y) \to (x, \lambda y) \mid \lambda^2=1 \} \mbox{ and } 
	\Gamma \S_\fbar = \{ \Id \}.
$$
They have the same dimension but  different cardinal. 
\end{remark}
		
\subsection{Expression swell}

In practice,  the above brute--force method, which consists 
of applying \RG\, to the PDE system \pref{PDEsystem},  is rarely
effective due to expressions swell. Much of the examples treated 
here and in \cite{dridi:these}, using our  algorithm {\sf ChgtCoords},
can not be treated with this approach.

It seems that the problem lies in the fact that we can not separate
the computation of $C_\varphi[\fbar]$ from that of $C_f[\fbar]$ which contains,
very often, big expressions. 

An other disadvantage of the above method is that  we have to restart computation from the very beginning
if  the target equation is changed. 
In the  next section, we propose our algorithm {\sf ChgtCoords} 
to compute the transformation $\varphi$ alone and in terms of differential
invariants. These invariants are provided by Cartan method for a generic 
$f$ which means that we have not re-apply Cartan method if the target
equation is changed and a big part of calculations is generic. 
Furthermore, the computation of $\varphi$ in terms of differential invariants 
reduces significantly the size of the expressions.

\section{Using Cartan's method}\label{cartan}

In this paper, differential invariants are obtained using Cartan's equivalence method. 
We refer the reader to \cite{cartan:pb,neut:these, Olver1, Hsu2} for
an  expanded tutorial presentation and application to second order ODE.
When applied Cartan's  method  furnishes a finite set of fundamental invariants  and 
a certain number of invariant derivations generating the differential field of invariant
functions. 

\begin{example}
Consider the EPB $(J^1(\C, \C),  \Phi_3)$. The
PDE system \pref{PDEsystem} reads
$$
\underbrace{
\begin{pmatrix}
\d\bar p-\fbar(\bar x,\bar y,\bar p)\d\bar x\\
\d\bar y-\bar p\d\bar x\\
\d\bar x
\end{pmatrix}}_{\omega_\fbar}
=
\underbrace{
\begin{pmatrix}
a_1 & a_2 & 0\\
0 & a_3 & 0\\
0 & 0 & 1\\ 
\end{pmatrix}}_{S(\a)}
\underbrace{
\begin{pmatrix}
\d p- f(x, y,p)\d x\\
\d y- p\d x\\
\d x
\end{pmatrix}}_{\omega_f}
$$
with $\det(S(\a))\neq 0$. In accordance with Cartan, this system is lifted
to the linear Pfaffian system 
$$
	S(\bar \a)\ \omega_\fbar=S(\a)\ \omega_f
$$ 
defined on the manifold of local coordinates $(\x,\a,\bar \x,\bar \a)$. 
After two normalizations and one prolongation, Cartan's method yields 
three fundamental invariants ($p=y'$ and $a=a_3$)
\begin{equation} \label{invs}
\begin{array}{ll}
  I_1 = -\Frac{1}{4}(f_{p})^{2} - f_{y} + \Frac{1}{2}D_{x}f_{p}, &
  I_2 = \Frac{f_{ppp}}{2 {a}^{2}},  \\[3mm]
  I_3 = \Frac{f_{yp} - D_{x}f_{pp}}{2a},
\end{array}
\end{equation}%
and the invariant derivations
\begin{equation}\label{derivations}
\begin{array}{l}
	X_1 = \Frac{1}{a}\Frac{\partial}{\partial p}, \quad
	X_3 = D_x- \Frac{1}{2}f_{p}a\Frac{\partial}{\partial a}, \quad
	X_4 = a\Frac{\partial}{\partial a},\\[3mm]
	X_2 =  \Frac{1}{a}\Frac{\partial}{\partial y} 
    		+ \Frac{1}{2}\Frac{f_{p}}{a}\Frac{\partial}{\partial p} 
    		- \Frac{1}{2}f_{pp}\Frac{\partial}{\partial a},
\end{array}
\end{equation}
where $D_x=\frac{\partial }{\partial x} +  p\frac{\partial }{\partial y} 
          + f(x,y,p)\frac{\partial }{\partial p}$. 

When $\dim(\S_f)=0$, the additional parameter $a$
can be (post)normalized by fixing some invariant to some suitable value. 
In this manner one constructs invariants  defined on~$M$ (not depending 
on the additional parameter).          
\end{example}

\begin{theorem}[Olver \cite{Olver1}]\label{tmOlver}
	If $\dim(\S_\fbar)=0$, then there exist exactly 
	$m$ functionally independent specialized invariants 
	$I_1[\fbar], \cdots, I_m[\fbar]$. 
\end{theorem}

Note that the  invariants  $I_1[\fbar], \cdots, I_m[\fbar]$ are functionally
independent if and only if  $\d I_1[\fbar]\wedge \cdots\wedge
\d I_m[\fbar] \neq 0$.
Note also that if  the function $\fbar$ is rational,  then 
the specialized invariants $I[\fbar]: M \to \C$ are algebraic functions.
In the sequel, we use the notation
$I_{i;j\cdots k}$ to denote the differential invariant $X_k\cdots X_j(I_i)$.
\subsection{Computation of $\varphi$}

Suppose that   the $\D$-groupoid $\S_\fbar$ is zerodimensional. Then, according to 
the theorem \ref{tmOlver}, there exists $m$ functionally independent invariants
$F_k := I_k[\fbar],$  $1 \leq k \leq m$. This implies that the 
 algebraic (non differential) system
\begin{equation}\label{sys:direct}
    \left\{ F_1(\xbar) =  I_1,\dots,  F_{m}(\xbar) = I_m \right\}
\end{equation}
is locally invertible and has a finite number of solutions 
\begin{equation} \label{sys:ivrs}
	\xbar = F^{-1}(I_1,\ldots, I_m). 
\end{equation}
The specialization of $I_1, \ldots , I_m$ on the source function  $f$ yields
\begin{equation}\label{cartan:phi}
	\bar \x= F^{-1}(I_1[f],\,\ldots, I_m[f]).
\end{equation}

Let $C$ denote the (non differential) characteristic set associated to the 
 system \pref{sys:direct} w.r.t. the elimination ranking 
$\{ \bar x,\, \bar y,\, \ldots, \bar y_n \} \succ \{I_1,\, \ldots,I_m \}$. Thus, $C$
describes the inversion \pref{sys:ivrs}.
The most simple situation happens when $\deg(C)=1$.
In this case, the necessary form of the change of coordinates 
$\varphi$ is the rational transformation defined by~$C$. 

\begin{example}
Consider the EPB $(\J^1(\C,\C), \Phi_3)$ and the target equation $\E_\fbar$
introduced by G. Reid \cite{Reid93}
$$
	\bar y''=\Frac{\bar y'}{\bar x} + \Frac{4\bar y^2}{\bar x^3}.
$$  
The following invariants are functionally independent
$$
\begin{gathered}
\bar I_{1;23} =-20{\Frac {1}{\bar a{\bar x}^{4}}},\ \bar I_{1;31}=8{\Frac {1}{\bar a{\bar x}^{3}}},\ 
 \bar I_{1}=\Frac{3}{4\bar x^2} +8\,{\Frac {\bar y}{{\bar x}^{3}}},\\
 \bar I_{1;3} ={\Frac {-3\,\bar x-48\,\bar y+16\,\bar p\bar x}{2{\bar x}^{4}}}.
\end{gathered}
$$
We normalize the parameter $\bar a$ by setting $\bar I_{1,23}=-20$. 
The characteristic set $C$ is
\begin{system}\nonumber
\bar p  =-{\Frac {3}{32}}+{\Frac {3}{512}} {I_{1;31  }}^{2}I_1+{\Frac {1}{4096}} 
I_{1;3  }{I_{1;31  }}^{3},\\[5mm]\bar y  =-{\Frac {3}{256}} I_{1;31  }+{\Frac 
{1}{4096}} I_{I_{1;31 }}^{3},\\[5mm]\bar x  =\Frac{1}{8} I_{1;31 },
\end{system}%
which gives the sought necessary form of $\varphi$. As a byproduct we deduce that the symmetry
group $\Gamma \S_\fbar = \{ \Id \}$. 
\end{example}

Let us return to the general situation, that is when  $\deg(C)$ is strictly bigger than 1. 
We have two cases.  First, $\deg(C)=\deg(\S_\fbar)$ and then $\varphi$ is the algebraic transformation defined
by $C$. Second,    $\deg(C) > \deg(\S_\fbar)$. 
In this  case, to obtain the transformation $\varphi$, we have to look for~$m$ other
functionally independent  invariants such that the new characteristic set~$C$ has
degree equal to $\deg(\S_\fbar)$.

\begin{example}
Consider the EPB $(\J^1(\C,\C), \Phi_3)$ and the target equation $\E_\fbar$
(number 8 in  \cite{Kamke})
$$
	\bar y''=\bar y^3 + \bar x\bar y
$$
which the corresponding  symmetry group  is 
$$
	\Gamma \S_\fbar =\left\{(x,y) \into (x,\,\lambda y) \, |
          \lambda^2=1 \right\}.
$$

One can verify that  $I_1,\, I_{1;13}$ and $I_{1;133}$,
when specialized on the considered equation, are functionally independent. 
In this case, the associated  characteristic set $C$~is
\begin{system}\nonumber
\bar p &=&- {\Frac { \left(4 {\bar x}^{2}+ 2 {I_1}\bar x-3 {I_{1;33}}
        -2 {I_1}^{2} \right) \bar y}{ 3(I_{1;3}+1)}},\\
{\bar y}^{2}&=& - \Frac{1}{3}\bar x-\Frac{1}{3} {I_1},\\[3mm]
{\bar x}^{3} &=&  -\Frac{3}{2}{I_1} {\bar x }^{2}  +\Frac{3}{4} {I_{1;33}} \bar x 
              - \Frac{3}{4}{ I_{1;3}}-\Frac{3}{8} {{ I_{1;3}}}^{2}+ \Frac{3}{4}{I_{1;33}} {I_1}\\
             && + \Frac{1}{2} {{I_1}}^{3}-\Frac{3}{8}.
\end{system}%
The degree of this set is equal to 6 which is  different from the degree of the symmetry
groupoid.

However, if instead of the above invariants we consider  the invariants  
$K_1 := {I_{1;233}}/{I_{1;31}}$, $K_2 :={I_{1;234}}/{I_{1;31}}$ and
$K_3 := {I_{1:231}}/{I^2_{1;31}}$, we obtain 
\begin{system} \nonumber
	\bar p &=& -K_1\bar y,\\
	{\bar y}^{2} &=& \Frac{1}{6}K_3,\\
	\bar x &=& - \Frac{1}{6}K_3 +K_1.
\end{system}%
This characteristic  gives  the necessary form of $\varphi$ since  it  has  degree two.

\end{example}

\subsubsection{Heuristic of degree reduction}
In practice, 
 one has to  search the invariants  giving the required degree in the
algebra of invariants. However,  this is
 not  an easy task since this algebra can be very large  (although it is  algorithmic).
For this reason we provide an  important heuristic which enables us to 
obtain the desired invariants.
This heuristic is explained in the following example.   

\begin{example}
Consider the Emden Fowler equation \pref{emf:eq} and the $\D$-groupoid of
transformations $\Phi_3$. We have already computed the corresponding symmetry
groupoid. The specialization of the invariants  $I_1,\, I_{1;13}$ and $I_{1;133}$ gives three 
functionally independent functions. As explained
above,  we obtain the following characteristic set computed w.r.t. the ranking
$\bar p \succ \bar y \succ \bar x \succ  I_1 \succ  I_{1;3} \succ
I_{1;33}$
\begin{system}\label{3resG1}
\bar p &=& \left( \Frac{3}{8}I_{1}-
\Frac{1}{4}{\Frac {I_{1;33}}{I_{1}}}+ \Frac{1}{3} {\Frac
{{I_{1;3}}^{2}}{{I_{1}}^{2}}} \right)\bar x\bar y - \Frac{1}{6} {\Frac 
{I_{1;3}}{I_{1}}}\bar y,  \\\bar y^3 &=& \left( -\Frac{9}{4}-2 {\Frac {{{\it 
I_{13}}}^{2}}{{I_{1}}^{3}}}+ \Frac{3}{2} {\Frac {I_{1;33}}{{I_{1}}^{2}}} \right) 
\bar x-{\Frac {I_{1;3}}{{I_{1}}^{2}}},   \\\bar x^2 &=& 4 \left( {\Frac {I_{1;3} 
I_{1} }{9 {{\it I_1}}^{3}-8 {I_{1;3}}^{2}+6 I_{1;33} I_{1}}} \right)\bar x \\&& 
+ 8 {\Frac {{I_{1}}^{2}}{9 {I_{1}}^{3}-8 {I_{1;3}}^{2}+6 I_{1;33} I_{1}}}. 
\end{system}%
Comparing with the $\D$-groupoid of symmetries \pref{symemf} we deduce that, in
contrary to $\bar y$, the degree of $\bar x$ must be reduced to one.
This can be done  in the following manner. First, observe that the Lie defining
equations of $\Phi_3$,  more exactly $\bar x_p=0$, implies that~${X_1(\bar x)=0}$
where $X_1=\frac{\partial}{a\partial p}$ is the invariant derivation~\pref{derivations}. Now,
differentiate the last equation of the  characteristic set, which we write as
$\bar x^2=A\bar x+B$, w.r.t the derivation $X_1$. We find $A_{;1}\bar x +
B_{;1}=0$.
The coefficient of $\bar x$ in this equation, which is
invariant, could not vanish (since  it is  not identically zero when specializing on
the Emden--Fowler equation). Thus, 
$\bar x = - \Frac{  B_{;1} }{ A_{;1}  }$
or explicitly
\begin{equation}\label{xbar}
	\bar x= -2 {\Frac { K I_{1;1}  + I_1K_{;1}  }{ K I_{1;31} +{\it I_{1;3}}
  	K_{;1}  }}\ \mbox{\ with\ } K={\Frac {{{\it I_1}}}{9 {{\it I_1}}^{3}-8 {{\it
	I_{1;3}}}^{2}+6 {\it I_{1;33}} {\it I_1}}}.
\end{equation}
The necessary form of
the change of coordinates $\varphi$ is then given  by 
 \pref{xbar} and 
the two first equations of
\pref{3resG1}.
\end{example}
The above reasoning can be summarized as follows 

\begin{center}
\begin{tabular}{|p{8cm}| }
  \hline
   {\sc Procedure} {\sf ChgtCoords} \\
   {\bf Input :} $\E_\fbar$ and $\Phi$ such that $\dim(\S_\fbar)=0$ \\
   {\bf Output :} $\bar \x=\varphi(\x)$ the necessary form of the change of coordinates \\
   1- Find $m$ functionally independent invariants $(I_1[\fbar],\, \ldots,\, I_m[\fbar])$
    defined on $M$.\\
   2- Compute a char. set $C$ of the algebraic system \pref{sys:ivrs}. \\
   3- If $\deg(C)=1$ then Return $C$. \\
   4- Compute $\S_\fbar$ with \RG.\\
   5- WHILE $\deg(C) \neq \deg(\S_\fbar)$ DO \\
    ~~~~ Reduce the degree of $C$. \\
    END DO\\
   6- Return $C$.\\
  \hline
\end{tabular}
\end{center}

\section{The solver}

\subsection{Precalculation of $\varphi$}
\subsubsection{The first step : the adapted $\D$-groupoid}

Let $\Phi_1,\,\ldots,\, \Phi_7 \subset \J^\infty_{*}(\C^2,\C^2)$ denote  the $\D$-groupoids
 defined in  the 
table 1 above. It is not difficult to see that 
$\Phi_1\subset\Phi_3\subset\Phi_5$ and  
$\Phi_2\subset\Phi_4\subset\Phi_6$ and finally 
$\Phi_5, \Phi_6 \subset \Phi_7$.

Let $\d(\E_f,\,\Phi) := \dim\left(\aut(\E_f) \cap \Phi\right)$ where 
$\aut(\E_f)$ is the contact symmetry $\D$-groupoid of the second
order ODE~$\E_f$.
Let $\d_i := \d(\E_f,\Phi_i)$ for $1\leq i\leq 7$.
\begin{definition}[Signature]
	The \emph{signature} of  $\E_f$ is
	$$
	\sign(\E_f) :=\left( (\d_1, \d_3, \d_5),(\d_2, \d_4,\d_6), \d_7\right).
	$$
\end{definition}
Clearly, $(\d_1 \leq \d_3 \leq \d_5 \leq \d_7)$ and $(\d_2 \leq \d_4
\leq \d_6 \leq \d_7)$.  Recall  that the calculation of
theses dimensions  does not require  solving differential equations.
We shall say that the signature $\mathrm{sign}(\E_f)$ \emph{matches} the signature 
$\sign(\E_\fbar)$ if and only if
$\d_7=\bar \d_7$ and $(s_1=\bar s_1 \mbox{ or } s_2=\bar s_2)$ where $s_1$ and $s_2$ stand for
$(\d_1, \d_3, \d_5)$ and $(\d_2, \d_4,\d_6)$ resp. 
Two second order ODE $\E_f$ and $\E_\fbar$ are said to be \emph{strongly equivalent} if 
$$
	\exists \Phi\in\{ \Phi_1,\cdots, \Phi_7\}, 
	\exists \varphi\in \Phi,\, \varphi_*\E_f=\E_\fbar,\, \d(\E_\fbar,\Phi)=0.
$$
\begin{lemma}
If $\E_f$ and $\E_\fbar$ are strongly equivalent then their signatures
match. 
\end{lemma}
\begin{definition}[Adapted $\D$-groupoid] \label{def:adapted}
	A $\D$-groupoid $\Phi$ is said to be \emph{adapted} to the ODE $\E_f$ if $\d(\E_f,\, \Phi)=0$ and 
	$\Phi$ is maximal among  $\Phi_1,\,\cdots,\, \Phi_7$ satisfying this property.
\end{definition}

\begin{center}
{\scriptsize
\begin{tabular}{|c|p{3cm}|p{3.5cm}|}
\hline
&{Transformations}  & Equation number according to Kamke's book\\
\hline    $\Phi_1$
&$                \bar  x=x,\               \bar y=\eta(x,y),              $
&
1, 2, 4, 7, 10, 21, 23, 24, 30, 31, 32, 40, 42, 43, 45, 47, 50
  \\
\hline    $\Phi_3$ &$                \bar  x=x+C,\
         \bar y=\eta(x,y),              $
 &
11, 78, 79, 87, 90, 91, 92, 94, 97, 98, 105, 106, 156, 172
 \\
\hline
    $\Phi_5$ &$
               \bar  x=\xi(x),\
               \bar y=\eta(x,y),
              $
     & Null\\
\hline
    $\Phi_2$ &$ 
               \bar  x=\xi(x,y),\ 
               \bar y=y,
              $  
     &  81, 89, 133, 134, 135, 237 \\
\hline
    $\Phi_4$ &$ 
               \bar  x=\xi(x,y),\ 
               \bar y=y+C,
              $  
&  11, 79, 87, 90, 92, 93, 94, 97, 98, 99, 105, 106, 172, 178\\
\hline
    $\Phi_6$ &$ 
               \bar  x=\xi(x,y),\ 
               \bar y=\eta(y),
              $  
     & 80, 86, 156, 219, 233  \\
\hline
    $\Phi_7$ &$ 
               \bar  x=\xi(x,y),\ 
               \bar y=\eta(x,y),
              $  
     & 3, 5, 6, 8, 9, 27,  44, 52,  85, 95, 108, 142, 144, 145, 147, 171, 211, 212, 238\\
\hline
\end{tabular} \\
\label{table}
{\it Table 1.  }}
\end{center}
The above table associates to each equation in the third column its  adapted
groupoids. For instance, the first Painlev\'e equation (number 3)   appears 
in the last row which means that its adapted $\D$-groupoid is 
the point transformations $\D$-groupoid $\Phi_7$ . To the  Emden--Fowler  equation,  
number~11, we associate the $\D$-groupoids  $\Phi_3$  and $\Phi_4$. In the case 
of homogeneous linear second order ODE (e.g. Airy equation, Bessel equation, 
Gau\ss\ hyper-geometric equation) we prove that, generically, the 
adapted $\D$-groupoid is~$\Phi_4$.

\subsubsection{The second step}
Once the list of adapted $\D$-groupoids $\Phi$  is known, we proceed by computing the necessary form  of the change of 
coordinates $\varphi\in\Phi$ using {\sf ChgtCoords}. Doing so, we construct a \Maple\ table indexed by  Kamke's book 
equations and where  entries corresponding to the index $\E_\fbar$ are: 
\begin{marray}
 1- & the signature of $\E_\fbar$, \\
 2- & the list of the adapted $\D$-groupoids $\Phi$ of  $\E_\fbar$,\\ 
 3- & the necessary form of the change of coordinates $\varphi\in\Phi$.
\end{marray}
For instance, the entries associated to Rayleigh equation $y''+ {y'}^4 + y=0$ are: 
\begin{marray}
 1- & the signature $\left((0,1,1),(1,1,1),1\right)$, \\
 2- & the $\D$-groupoid~$\Phi_1$, \\
 3- & the necessary form of the change of coordinates 
\end{marray} 
\begin{system}\label{chgt2Ray}
	\bar p &=& -36{\Frac {I_{2;1}}{72+72I_1+
	{I_{2;1}}^{2}}}\bar y,\\
	\bar x &=&x,\\
	{\bar y}^{3} &=& {\Frac {-1}{559872 {I_{2;1}^{2}}} }(I_{2;1}^{6}+216I_1I_{2;1}^{4}
	+ 216I_{2;1}^{4} +373248
	\\
	&& +15552I_{2;1}^{2} +1119744I_1^{2}+  31104I_1I_{2;1}^{2}\\
	&&+1119744I_1+373248I_1^{3}+15552I_1^{2}I_{2;1}^{2})
\end{system}%
with the normalization 
$
{I_2}/{I_{2;1}}=1.
$
Invariants here are those generated by  \pref{invs} and \pref{derivations} plus 
the essential invariant $\bar x=x$.

\subsection{Algorithmic scheme of the solver}
To integrate a differential equation $\E_f$ our solver proceeds as follows
\begin{center}
\begin{tabular}{|p{8cm}| }
  \hline
   {\sc Procedure} {\sf Newdsolve}  \\
   {\bf Input :} $\E_f$  \\
   {\bf Output :} An equation $\E_\fbar$ in  Kamke's book and the transformation 
   $\varphi$ such that $\varphi_*(\E_f)=\E_\fbar$ \\
   1- Compute the signature of $\E_f$.\\
   2- Select from the table the list of equations $\E_\fbar$ such that $\sign(\E_\fbar)$
   matches $\sign(\E_f)$.\\
   3- FOR each equation $\E_\fbar$ in the selected list DO \\
   	~~(i) Specialize, on $\E_f$, the necessary form of the change of coordinates 
   	associated to $\E_\fbar$. We obtain~$\varphi$.\\
   	~~(ii) If $\varphi \in \Phi$ and $\varphi_*(\E_f)=\E_\fbar$ 
     	then return ($\E_\fbar$, $\varphi$). \\
    END DO. \\
  \hline
\end{tabular}
\end{center}

It is worth noticing  that the  time required to perform steps (i)- (ii) is very small. In fact, it is 
about one hundredth of a  second using  Pentium(4) with 256 Mo.
 The second feature of our solver is,  contrarily  to the symmetry methods, neither the table 
construction nor the algorithm of the solver involves integration
of differential equations.

\section{Acknowledgments}
We are thankful to Rudolf Bkouche and François Boulier for many useful discussions during the preparation of this article.


\def\cprime{$'$}

%
%

\balancecolumns 
\end{document}